# Escape dynamics of a particle from a purely-nonlinear truncated quartic potential well under harmonic excitation

Maor Farid

*Massachusetts Institute of Technology, 77 Massachusetts Ave., Cambridge, MA 02139, United States*
*Faculty of Mechanical Engineering, Technion – Israel Institute of Technology, Haifa 3200003, Israel*
*faridm@mit.edu*

## Abstract

This paper focuses on the escape problem of a harmonically-forced classical particle from a purely-quartic truncated potential well. The latter corresponds to various engineering systems that involve purely cubic restoring force and absence of linear stiffness even under the assumption of small oscillations, such as pre-tensioned metal wires and springs, and compliant structural components made of polymer materials. This, in contrast to previous studies where the equivalent potential well could be treated as linear at first approximation under the assumption of small perturbations. Due to the strong nonlinearity of the current potential well, traditional analytical methods are inapplicable for describing the transient bounded and escape dynamics of the particle. The latter is analyzed in the framework of isolated resonance approximation by canonical transformation to action-angle (AA) variables and the corresponding reduced resonance manifold (RM). The escape envelope is formulated analytically. Surprisingly, despite the essential nonlinearity of the well investigated, it exhibits a universal property of a sharp minimum due to the existence of multiple intersecting escape mechanisms. Unlike previous studies, three underlying mechanisms that govern the transient dynamics of the particle were identified: two maximum mechanisms and a saddle mechanism. The first two correspond to a gradual increase in the system's response amplitude for a proportional increase in the excitation intensity, and the latter corresponds to an abrupt increase in the system's response and therefore more potentially hazardous. The response of the particle is described in terms of energy-based response curves. The maximal transient energy is predicted analytically over the space of excitation parameters and described using iso-energy contours. All theoretical predictions are in complete agreement with numerical results.

## 1. Introduction

The profound physical problem of the escape of a classical particle from a potential well has numerous applications in classical physics, engineering, thermodynamics, and chemistry [1, 2, 3]. Escape might take place due to external disturbances of various kinds, such as non-zero initial conditions, impact, periodic or stochastic excitations. For periodic excitation, the curve that represents the critical excitation parameters associated with escape is referred to as the escape envelope. Previous works reported on a universal V-shaped escape curve whose minimum is located near the natural frequency of the particle [4, 5, 6]. This universal pattern stems from the co-existence of two competing escape mechanisms that can be described using the assumption of 1:1 isolated resonance, allowing an efficient order reducing if the system's dynamics onto a 2D phase portrait of a 1:1 resonance manifold (RM). The trajectory which corresponds to zero initial conditions, and therefore relevant for the current analysis, is called the limiting phase trajectory (LPT) [7, 8]. In the phase portrait of the system, escape corresponds to the continuous passage of the LPT from zero initial energy to the critical energy level associated with escape. The first escape mechanism is referred to as the 'maximum mechanism' (MM). In this scenario, the LPT approaches directly from the bottom of the phase portrait to its upper bound, and therefore, from the well's equilibrium point to its top edge. The second scenario is called the 'saddle mechanism' (SM) since it corresponds to the passage of the LPT through the saddle point of the RM followed by a sudden jump towards the upper bound of the well. In this case, the particle's displacement increases abruptly, which makes this mechanism much more potentially hazardous when takes place in engineering systems. On the other hand, in previous works it was demonstrated that this mechanism is highly effective for energy absorption and can be utilized by nonlinear energy sinks (NESs) for efficient vibration mitigation purposes [9].

Previous studies mainly focused on transient escape dynamics of weakly-nonlinear potential wells [10, 5, 6, 11]. There, the nonlinear features of the well could usually be omitted at first approximation under the assumption of small oscillations. In the current work, we explore analytically the effect of the *absence* of a linear term in the potential well, as well as the effect of strongly-nonlinear quartic term and its effect on the escape mechanisms. Quartic potential corresponds to a cubic restoring force which is the most common nonlinearity in mechanical components and engineering systems that exhibit symmetry. Cubic stiffness can stem either from the system's structural configuration or from its material properties. Examples include pre-tensioned wires, springs, polymers, and foams [12]. Another engineering use-case of a cubic nonlinearity is passive energy absorption; the cubic NES utilizes its purely-cubic nonlinearity in the form of an adaptive-frequency that allows it to exhibit good vibration mitigation capabilities for a broad frequency range [13, 14].

While previous studies focused merely on the particle's escape from the well under harmonic excitation, in the current study we focus on both its bounded and escape dynamics. Two types of bifurcations are introduced: bifurcation of type I is referred to the escape from the well, while bifurcation of type II corresponds to reaching a critical maximal energy level. In other terms, the former is a particular case of the latter for the critical energy level of one. We use this distinction to describe both the escape envelope of the potential well over the space of excitation parameters (bifurcation of type I) and other contour lines associated with lower maximal transient energy levels. Those contours accumulate to a manifold over the excitation parameters space that maps sets of excitation parameters to their corresponding predicted maximal transient energy levels.

This paper is structured as follows: In Section 2 the dynamical model of the classical particle in the truncated quartic potential well is introduced. In Section 3, the dynamical response of the bounded particle is described analytically using an action-angle formalism in the perspective of the topology of an underlying resonance manifold. In Section 4, the maximal transient energy for any given set of excitation parameters is predicted analytically, and in Section 5 the escape envelope of the particle is constructed on the space of excitation parameters. In Section 6, the energy-based frequency response curve of the forced particle is obtained. In Section 7, all analytical results are validated numerically. Finally, Section 8 is dedicated to concluding remarks.

## 2. Model description

In order to capture the dynamical effect of a purely-cubic restoring force, the following truncated quartic potential well is adopted, where $q$ is a non-dimensional displacement variable. A sketch of the potential well is shown in Fig. 1.

$$U(q) = \begin{cases} q^4 & , |q| < 1 \\ 1 & , \text{else} \end{cases} \qquad (1)$$

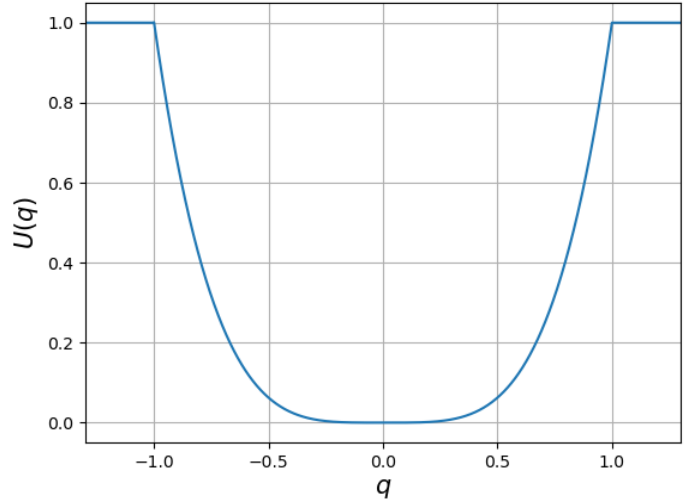

Figure 1: The equivalent quartic potential well. Corresponds to a dynamical system with a purely-cubic stiffness. Energy values of $U(q) \in (0,1)$ correspond to bounded dynamics and values of $U(q) \geq 1$ correspond to escape from the well.

The non-dimensional equation of motion of the particle under a periodic mono-chromatic excitation is derived from Eq. (1) as follows:

$$\ddot{q} + 4q^3 = f\cos(\Omega \tau) \qquad (2)$$

Here $\Omega$ and $f$ are the non-dimensional frequency and amplitude of the external excitation and $\tau$ is a non-dimensional time scale. The particle escapes from the well when it reaches the upper bound of the well, i.e. when $U(q) = 1$. Hence, the following displacement-based escape criterion is adopted:

$$\max_\tau |q(\tau)| = 1 \qquad (3)$$

## 3. Topology exploration of the resonance manifold

The dynamical system in Eq. (2) can be derived from the following underlying Hamiltonian:

$$H = H_0(p,q) - fq\sin(\Omega\tau), \quad H_0(p,q) = \frac{1}{2}p^2 + q^4 \qquad (4)$$

Here, $H_0 = E$ is the conservative component of the Hamiltonian, and $E$ is the energy of the particle. It corresponds to the free motion of the particle in absence of external excitation and is determined only by initial conditions. The momentum of the particle is denoted by $p = \dot{q}$. The canonical transformation to action-angle (AA) variables is performed using the following well-known formulas:

$$I(E) = \frac{1}{2\pi}\oint p(q,E)dq, \quad \theta = \frac{\partial}{\partial I}\int_0^q p(q,I)dq \qquad (5)$$

Here $I$ and $\theta$ are the action and angle variables, respectively. Theoretically, expressions

$p(I, \theta)$ and $q(I, \theta)$ can be obtained by inverting the relations in Eq. (5). The conservative component of the Hamiltonian depends only on the action variable: $H_0 = E(I)$. Now, the Hamiltonian in Eq. (4) can be rewritten in terms of AA variables as follows:

$$H = H_0(I) - fq(I, \theta) \sin(\Omega \tau) \tag{6}$$

Due to the periodicity of the second term in Eq. (6) it can be expressed using a Fourier series as follows [15]:

$$H = H_0(I) + \frac{if}{2} \sum_{n=-\infty}^{\infty} q_n(I) \left( e^{i(n\theta + \Omega\tau)} - e^{-i(n\theta - \Omega\tau)} \right), \quad q_n = \bar{q}_{-n} \tag{7}$$

Here bar stands for a complex conjugate. The following well-known relations between the action and angle variables and the Hamiltonian are used:

$$\dot{I} = -\frac{\partial H}{\partial \theta}, \quad \dot{\theta} = \frac{\partial H}{\partial I} \tag{8}$$

Following Eq. (7) and Eq. (8), the Hamilton equations take the following form:

$$\begin{aligned}
\dot{I} &= \frac{f}{2} \sum_{n=-\infty}^{\infty} n q_n(I) \left( e^{i(m\theta + \Omega\tau)} - e^{i(m\theta - \Omega\tau)} \right) \\
\dot{\theta} &= \frac{\partial H_0}{\partial I} + \frac{if}{2} \sum_{n=-\infty}^{\infty} \frac{\partial q_n(I)}{\partial I} \left( e^{i(m\theta + \Omega\tau)} - e^{i(m\theta - \Omega\tau)} \right) - \Omega
\end{aligned} \tag{9}$$

In the current analysis, we assume that the excitation frequency is in the vicinity of one, i.e. $\Omega \approx 1$. In order to analyze this case, we also assume slow evolution of the phase variable $\nu = \theta - \Omega\tau$ with respect to all the other phase combinations in Eq. (9). Hence, averaging over these fast time scales yields the following slow-flow equations:

$$\begin{aligned}
\dot{J} &= -\frac{f}{2} \left( q_1(J) e^{i\nu} + \bar{q}_1(J) e^{-i\nu} \right) \\
\dot{\nu} &= \frac{\partial H_0}{\partial J} - \frac{if}{2} \left( \frac{\partial q_1(J)}{\partial J} e^{i\nu} - \frac{\partial \bar{q}_1(J)}{\partial J} e^{-i\nu} \right) - \Omega
\end{aligned} \tag{10}$$

Here $J(\tau) = \langle I(\tau) \rangle$ is the averaged action variable and variable $q_1(J)$ is the first Fourier coefficient of $q(I, \theta)$. By direct integration of the system in Eq. (10), the following conservation rule is obtained:

$$C(J, \nu) = H_0(J) - \frac{if}{2} \left( q_1(J) e^{i\nu} - \bar{q}_1(J) e^{-i\nu} \right) - \Omega J \tag{11}$$

Eq. (11) represents a family of resonance manifolds (RMs), where variable $C$ is determined by initial conditions. In the current study, we consider the case in which the particle begins its motion from rest at the bottom of the well, i.e. with zero initial conditions. The contour on the RM which is associated with this case is referred to as the limiting phase trajectory (LPT) and corresponds to $H_0(J) = 0$ [6]. From Eq. (4)-(5), the averaged action variable is calculated as follows:

$$J(\xi) = \beta \xi^{\frac{3}{4}}, \quad \beta = \frac{4}{3\pi} \mathbf{K}\left(\frac{1}{\sqrt{2}}\right) \tag{12}$$

Here, the averaged energy of the particle is denoted by $\xi(\tau) = \langle E(\tau) \rangle$ and $\mathbf{K}$ is the elliptic integral of the first kind. The frequency of the particle's response is obtained as follows:

$$\omega(\xi) = \left(\frac{\partial J}{\partial \xi}\right)^{-1} = \frac{\pi}{\mathbf{K}\left(\frac{1}{\sqrt{2}}\right)} \xi^{\frac{1}{4}} \qquad (13)$$

In Fig. 2, the monotonic increase of the response frequency $\omega(\xi)$ stems from hardening nonlinearity associated with the positive cubic term in the equation of motion of the system (Eq. (2)) [16, 17].

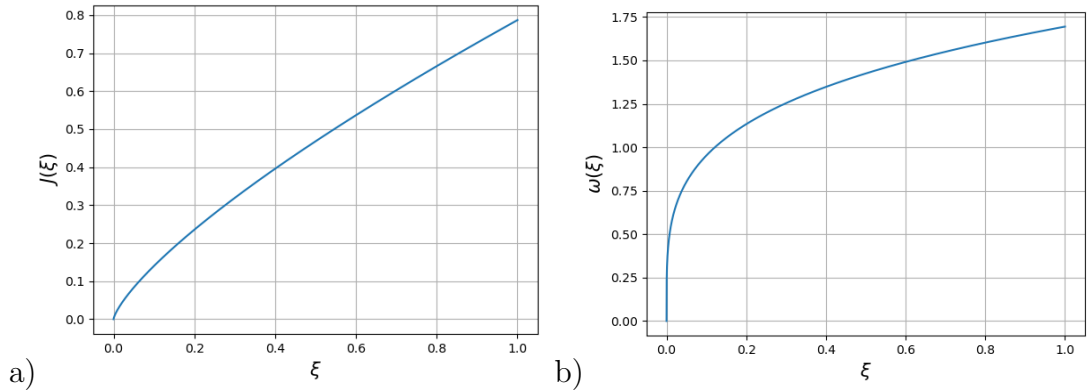

Figure 2: a) The averaged action, and b) the response frequency of the particle vs. its averaged energy $\xi$, according to Eq. (12) and Eq. (13), respectively.

Following Eq. (5), the solution of the system is given in terms of AA variables in the following form:

$$q(\theta, \xi) = \xi^{\frac{1}{4}} \mathrm{sn}(\theta, i) \qquad (14)$$

Here $\mathrm{sn}(x, i)$ is the Jacobi elliptic function of module $i = \sqrt{-1}$. Recalling Eq. (11), the particle's displacement can be approximated using the following Fourier series:

$$q(\theta, \xi) = a_0 + \sum_{n=1}^{\infty} a_n \sin(n\theta) \qquad (15)$$

Detailed derivation of the AA transformation is shown in Appendix A. By following Eq. (11), Eq. (14), and Eq. (15), and recalling that $q_1 = -ia_1/2$, we calculate the coefficient of the first term of the Fourier series, i.e. $a_1$. The latter is given in Eq. (16) and plotted in Fig. 3.

$$a_1(\xi) = \alpha \xi^{\frac{1}{4}}, \qquad \alpha = \frac{2\pi\eta}{\mathbf{K}(i)(1+\eta^2)} \qquad (16)$$

Here, $\eta = e^{-\pi/2}$. Following [6], all values in the current analysis will be expressed in terms of the averaged energy $\xi$ and not the averaged action $J$. Consequently, the conservation law gets the following form:

$$C(\nu, \xi) = \xi - \frac{f}{2} a_1(\xi) \cos(\nu) - \Omega J \qquad (17)$$

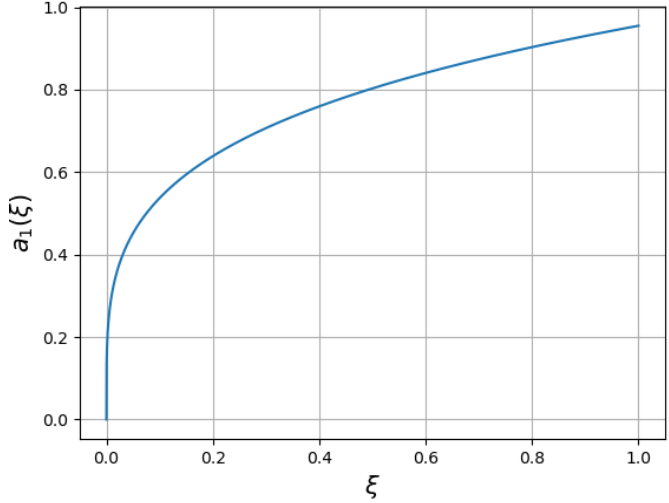

Figure 3: The first Fourier coefficient of $q(\theta,\xi)$ vs. the averaged energy of a particle in a quartic potential well , $a_1(\xi)$.

As mentioned above, the current analysis focuses on the response of the system for zero initial conditions. The latter is represented by the zero level line of the RM, called the LPT, that corresponds to $C(\nu,\xi) = 0$. The maximal energy level reached by the continuous LPT corresponds to the maximal energy absorbed by the particle. The geometry of the LPT is dictated by the topology of the RM that deforms with changing the excitation parameters. The topological structure and phase portrait of the RM are defined by the conservation law in Eq. (17). The stationary points of the RM correspond to the following equations:

$$\frac{\partial C(\nu,\xi)}{\partial \nu} = 0, \quad \frac{\partial C(\nu,\xi)}{\partial \xi} = 0 \qquad (18)$$

Solving the first equation in Eq. (18) yields that the stationary points of the RM are located on lines $\nu = 0, \pi$ on the phase cylinder. The second equation yields the relations between forcing parameters and the location of the saddle point of the RM at $\nu = \pi$, as described in the next section.

## 4. Maximal transient energy

The main goal of this section is to obtain an analytical prediction of the transient maximal energy reached by the particle for any given set of excitation parameters $\Omega$ and $f$. In engineering systems, the transient energy can serve as a measure of the vibration-induced damage. On the other hand, in PEAs the transient energy absorption rate usually serves as a measure for the vibration mitigation capabilities of the PEA [18, 19, 20]. The maximal transient energy is obtained through several underlying dynamical mechanisms that correspond to the changing topology of the RM. For the sake of clarity, let us distinguish between two scenarios: the first is an escape from the well that corresponds to maximal transient energy of one $\tilde{\xi} = 1$. The second scenario corresponds to the case of maximal transient energy that reaches an energy level of interest $\tilde{\xi}$. Both mechanisms are referred to as bifurcations of type I and type II, respectively. It can be easily seen that the type I bifurcation is a particular case of the type II bifurcation for $\tilde{\xi} = 1$. In the current section, we focus on the dynamical mechanisms that govern the increase of the transient energy of the particle, while creating the analytical infrastructure for exploring the escape phenomenon that will be treated separately in Section 5.

## 4.1. Saddle mechanism

Passage of the LPT through the RM's stationary points at $\nu = \pi$ corresponds to the following conditions, obtained from Eq. (17)-(18):

$$\begin{cases} C(\nu = \pi, \xi_s | f_s) = \xi_s + \frac{f_s}{2} a_1(\xi_s) - \Omega J(\xi_s) = 0 \\ \frac{\partial C}{\partial \xi}(\nu = \pi, \xi_s | f_s) = 1 + \frac{f_s}{2} a_1'(\xi_s) - \Omega J'(\xi_s) = 0 \end{cases} \quad (19)$$

Here $\xi_s$ is the energy level of the stationary points of the RM at $\nu = \pi$ and $f_s$ is the critical excitation amplitude associated with the passage of the LPT through them. Tag represents differentiation with respect to the averaged energy $\xi$. The system of equations in Eq. (19) yields the following relation between the excitation frequency and the energy level of the single stationary points of the RM on line $\nu = \pi$:

$$\xi_s = \left(\frac{2\beta}{3}\right)^4 \Omega^4 \quad (20)$$

As one can learn by plotting $C(\nu, \xi)$, the single stationary point of the RM at $\nu = \pi$ is a saddle. The energy level of the saddle point $\xi_s$ vs. the forcing frequency $\Omega$ is plotted in Fig. 4. As one can see, the energy level of the saddle point increases monotonously with the forcing frequency. However, when $\xi_s$ reaches the upper boundary of the phase cylinder $(\nu, \xi) \in (0 - 2\pi, 0 - 1)$ the saddle point (and therefore the SM) ceases to exist. The critical forcing frequency associated with the disappearance of the saddle point is denoted by $\hat{\Omega}$. The alter is calculated in Eq. (21) and described by a vertical dashed line in Fig. 4.

$$\xi_s = 1 \rightarrow \hat{\Omega} = \frac{3}{2\beta} \approx 1.9062 \quad (21)$$

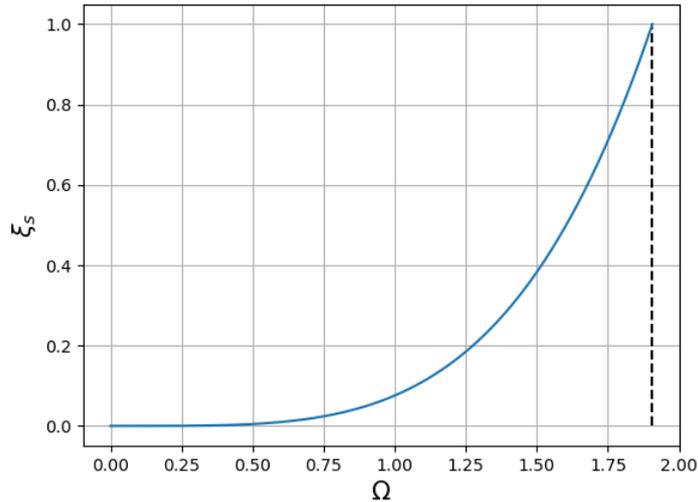

Figure 4: Plot of the energy level associated with the saddle point $\xi_s$ vs. the excitation frequency according to Eq. (20). The vertical black dashed line corresponds to the critical excitation frequency $\hat{\Omega} \approx 1.9062$ that corresponds to the disappearance of the saddle point (Eq. (21)).

Substituting Eq. (20) into system (19) yields the following expression of the critical excitation amplitude associated with bifurcation through the SM:

$$f_s(\Omega) = \frac{2}{a_1(\xi_s)}\left(\Omega J(\xi_s) - \xi_s\right) = \frac{8\beta^3}{27\alpha}\Omega^3 \qquad (22)$$

Substituting Eq. (22) into Eq. (20) yields the critical excitation amplitude associated with the disappearance of the saddle point and the SM:

$$\hat{f} = f_s(\hat{\Omega}) = \frac{1}{\alpha} \approx 1.0471 \qquad (23)$$

As one can learn from (22), the critical excitation amplitude associated with the SM does not depend on the critical transient energy level $\tilde{\xi}$. Bifurcation of type I (escape) thought the SM is demonstrated in Fig. 5 from the perspective of the phase portrait. There, both branches of the LPT intersect at $\xi_s$, creating a continuous connection between the bottom and top boundaries of the phase cylinder. Similar structure takes place in bifurcation of type II for $\tilde{\xi} < 1$.

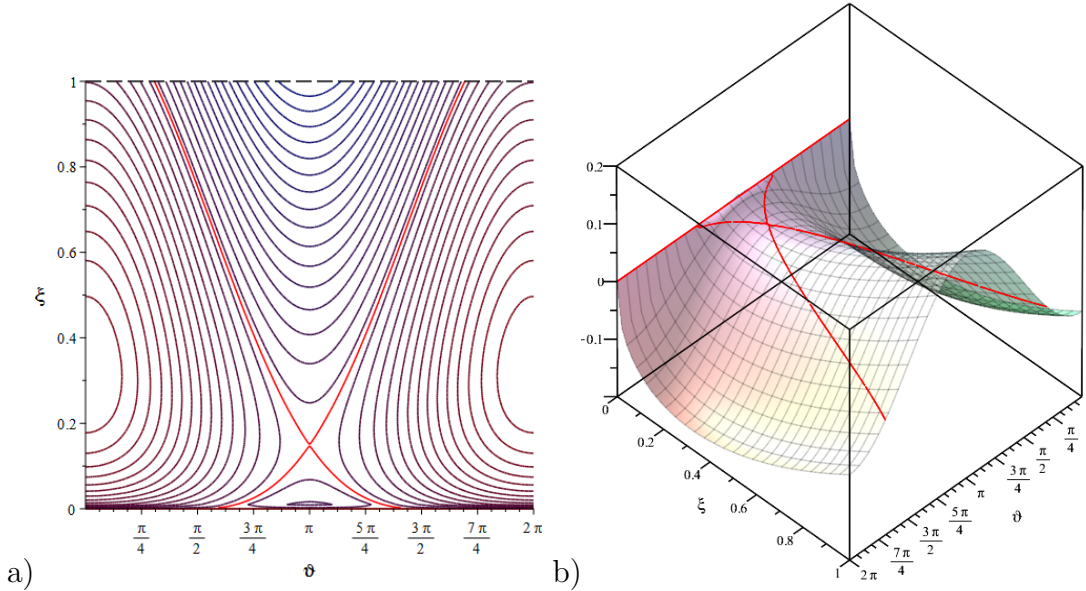

Figure 5: Type I bifurcation (escape) through the saddle mechanism in phase portrait, defined by $C(\nu, \xi)$. The LPT is marked by a red line. Dasehd black line correspond to the upper bound of the phase cylinder that corresponds to escape from the well, i.e. $\tilde{\xi} = 1$. Type II bifurcation has similar structure for $\tilde{\xi} < 1$. For $\Omega = 1.184$, and $f = 0.2509$, a) 2D projection of $C(\nu, \xi)$ on the $(\nu, \xi)$ plane, b) 3D plot of $C(\nu, \xi)$.

### 4.2. Maximum mechanism at $\nu = 0$

For $f > f_s(\Omega)$, bifurcations are dominated by the maximum mechanism (MM) at $\nu = 0$. In this scenario, the LPT directly approaches the critical energy level $\tilde{\xi}$ at $\nu = 0$ (bifurcation of type I: $\tilde{\xi} = 1$, bifurcation of type II: $\tilde{\xi} < 1$), without passing thought the saddle point at $\nu = \pi$. Thus, this MM is referred to as the MM0. The corresponding critical forcing amplitude is obtained from Eq. (18) using the following set of equations:

$$\begin{cases} C(\nu = 0, \tilde{\xi}|f_{m,0}) = \tilde{\xi} - \frac{f_{m,0}}{2}a_1(\tilde{\xi}) - \Omega J(\tilde{\xi}) = 0 \\ \frac{\partial C}{\partial \xi}(\nu = 0, \tilde{\xi}|f_{m,0}) = 1 - \frac{f_{m,0}}{2}\frac{\partial a_1(\tilde{\xi})}{\partial \xi} - \Omega \frac{\partial J(\tilde{\xi})}{\partial \xi} = 0 \end{cases} \qquad (24)$$

The relations in Eq. (24) correspond to the fact that the LPT is tangent to the upper bound of the phase cylinder $\tilde{\xi}$ at $\nu = 0$. Solving Eq. (24) yields the following relation between the excitation frequency to the critical forcing amplitude:

$$f_{m,0}(\Omega|\tilde{\xi}) = \frac{2}{a_1(\tilde{\xi})}\left(\tilde{\xi} - \Omega J(\tilde{\xi})\right) = \frac{2}{\alpha}\left(\tilde{\xi}^{\frac{3}{4}} - \beta\Omega\tilde{\xi}^{\frac{1}{2}}\right) \quad (25)$$

As mentioned above, the MM0 is dominant only above the curve described in Eq. (22), i.e. for $f > f_s(\Omega)$. Bifurcation of type I (escape) thought the MM0 is demonstrated graphically in Fig. 6 from the perspective of the phase portrait.

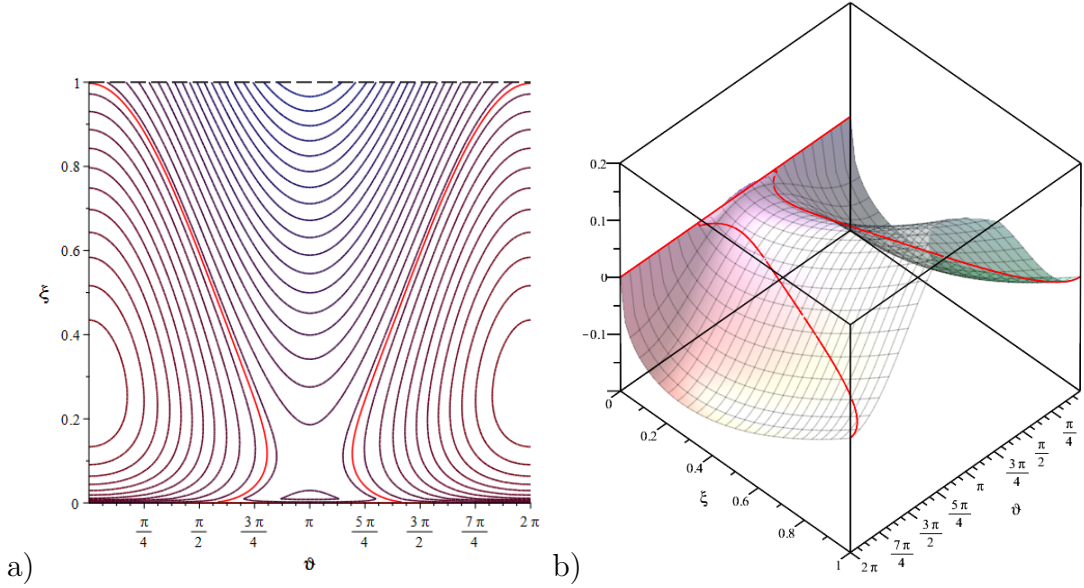

Figure 6: Type I bifurcation through the maximum mechanism at $\nu = 0$ (MM0) in the phase portrait, defined by $C(\nu, \xi)$. The LPT is marked by a red line. Dased black line correspond to the upper bound of the phase cylinder that corresponds to escape from the well, i.e. $\tilde{\xi} = 1$. For $\Omega = 1.115$, $f = 0.255$, and a) 2D projection of $C(\nu, \xi)$ on the $(\nu, \xi)$ plane, b) 3D plot of $C(\nu, \xi)$.

*4.3. Maximum mechanism at $\nu = \pi$*

In contrast to the SM, for $f < f_s(\Omega)$, the upper and the lower branches of the LPT are yet to intersect since the forcing amplitude is insufficient for saddle bifurcation to take place. In this case, a sub-saddle MM takes place below the saddle point at $\nu = \pi$, i.e. for $\tilde{\xi} < \xi_s(\Omega)$. This mechanism is referred to as the MM$\pi$, and it satisfies the following set of equations:

$$\begin{cases} C(\nu = \pi, \tilde{\xi}|f_{m,\pi}) = \tilde{\xi} + \frac{f_{m,\pi}}{2}a_1(\tilde{\xi}) - \Omega J(\tilde{\xi}) = 0 \\ \frac{\partial C}{\partial \xi}(\nu = \pi, \tilde{\xi}|f_{m,\pi}) = 1 + \frac{f_{m,\pi}}{2}\frac{\partial a_1(\tilde{\xi})}{\partial \xi} - \Omega\frac{\partial J(\tilde{\xi})}{\partial \xi} = 0 \end{cases} \quad (26)$$

The relations in Eq. (26) correspond to the fact that the LPT is tangent to the line $\tilde{\xi} < \xi_s(\Omega)$ on the phase cylinder at $\nu = \pi$. Solving Eq. (24) yields the following relation between the excitation frequency to the critical forcing amplitude:

$$f_{m,\pi}(\Omega|\tilde{\xi}) = -\frac{2}{a_1(\tilde{\xi})}\left(\tilde{\xi} - \Omega J(\tilde{\xi})\right) = -\frac{2}{\alpha}\left(\tilde{\xi}^{\frac{3}{4}} - \beta\Omega\tilde{\xi}^{\frac{1}{2}}\right) \quad (27)$$

Type I bifurcation (escape) thought the MM$\pi$ is demonstrated graphically in Fig. 7.

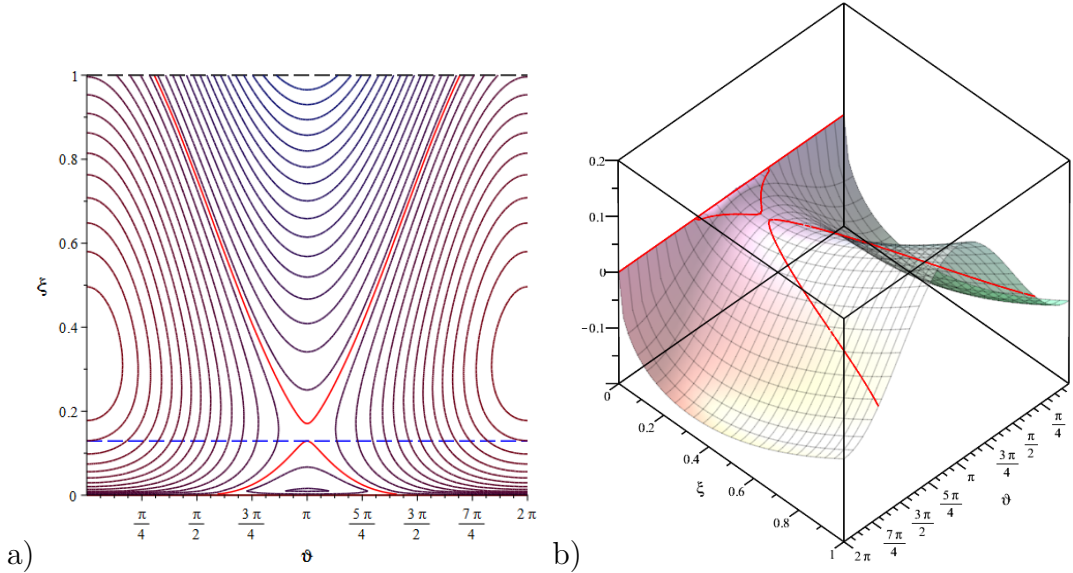

Figure 7: Escape through the maximum mechanism at $\nu = \pi$ (MM$\pi$) in the phase portrait, defined by $C(\nu, \xi)$. The LPT is marked by a red line. For $\Omega = 1.184, f = 0.25$. Dashed black and blue lines correspond to the upper bound of the phase cylinder ($\tilde{\xi} = 1$) and to the maximal energy level reached by the LPT, i.e. $\tilde{\xi} = 0.13$ (satisfies $\tilde{\xi} < \xi_s(\Omega = 1.184) = 0.1488$); a) 2D projection of $C(\nu, \xi)$ on the $(\nu, \xi)$ plane, b) 3D plot of $C(\nu, \xi)$.

Now, the maximal transient energy level reached for any set of $\Omega$ and $f$, i.e. $\tilde{\xi}(\Omega, f)$, can be obtained by inverting the relations in Eq.(25) and Eq.(27). Each of those cases corresponds to another region in the space of forcing parameters: the former (MM0) corresponds to $f > f_s(\Omega)$ and the latter (MM$\pi$) corresponds to $f < f_s(\Omega)$. Inversion of equations Eq. (25) and Eq. (27) yields the following pair of equations:

$$\tilde{\xi}^{\frac{3}{4}} - \beta\Omega\tilde{\xi}^{\frac{1}{2}} \pm \frac{\alpha}{2}f = 0 \rightarrow \tilde{\xi}(\Omega, f) \tag{28}$$

Here, the plus and minus signs corresponds to the ranges $f < f_s(\Omega)$ and $f > f_s(\Omega)$, respectively. The former yields two positive real (and thus physical) roots. The larger among them corresponds to $f > f_s(\Omega)$ and the latter to $f < f_s(\Omega)$ and therefore, the lower root is chosen. The latter yields a single positive real root. Recalling that the maximal transient energy level is bounded by the upper bound of the well $\tilde{\xi} = 1$, an injective mapping between excitation parameters and the resulting maximal energy level is obtained, as shown in Fig. 8. The black curve the divides the plane into two regions corresponds to $f_s(\Omega)$. The descending and ascending lines correspond to $f_{m,0}(\Omega|\tilde{\xi})$ ($f > f_s(\Omega)$) and $f_{m,\pi}(\Omega|\tilde{\xi})$ ($f < f_s(\Omega)$), respectively. The yellow region ($\tilde{\xi} = 1$) corresponds to escape from the well (bifurcation of type I). The black level lines are iso-energy lines that describe sets of excitation parameters that lead to identical maximal transient energy levels. The escape envelope of the well corresponds to the top iso-energy line of $\tilde{\xi} = 1$. As one can see in Fig. 8, the minimum of the iso-energy lines shifts to the right until the minimum of the escape curve is obtained in the vicinity of $\Omega = 1$. This similarity to weakly-nonlinear potential wells with a linear term is somewhat surprising. Although the current well is purely nonlinear and absent a linear term- the shape of the escape envelope and the location of its minimum is still preserved.

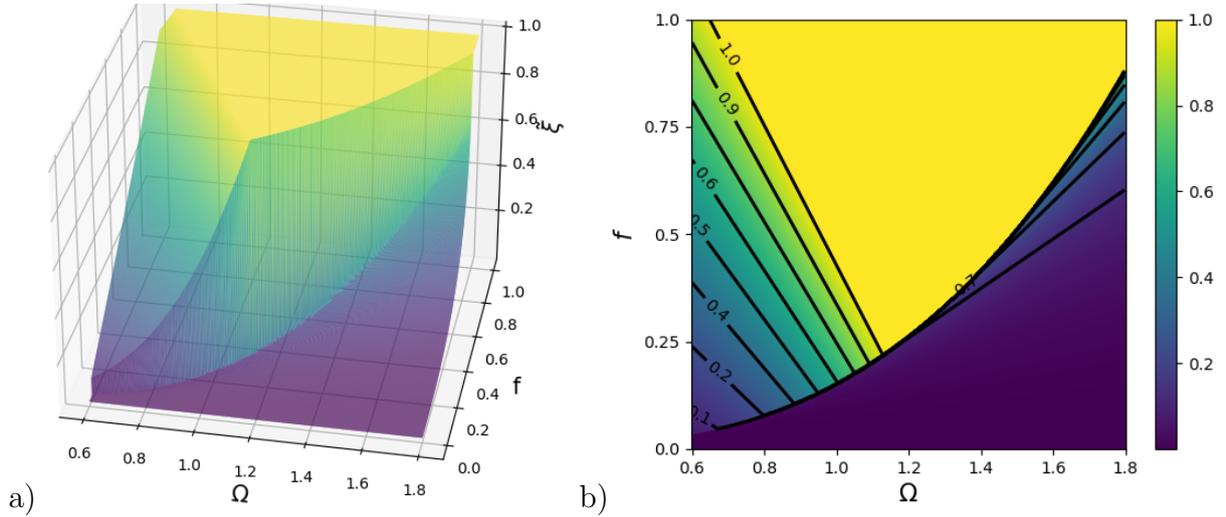

Figure 8: The maximal transient energy level over the plane of forcing parameters $\tilde{\xi}(\Omega, f)$; a) 3D plot, b) 2D projection with iso-energy level lines. Level line $\tilde{\xi} = 1$ corresponds to the escape curve of the potential well.

The graphical representation in Fig. 8 gives a full perspective on the predicted response energy levels for any set of excitation parameters. In the perspective of equivalent engineering systems or PEAs, Fig. 8 given the designer a complete understanding of the response of the system or energy absorption capabilities over the parameters space of monochromatic harmonic excitations.

## 5. Escape envelope

In the current section, the critical conditions for escape are discussed. The set of all critical excitation parameters that lead to escape are accumulated to a curve that is referred to as the escape envelope of the well. The latter separates the plane of excitation parameters into two basins: the escape basin, and the safe basin. The former is characterized by transient energy levels that exceed the critical threshold $\tilde{\xi} = 1$, while the latter is associated with lower energy levels $\tilde{\xi} < 1$ and correspond to the yellow basin in Fig. 8. The escape envelope of the well is the perimeter of the safe basin in the excitation parameters plane. The main goal of the current section is to obtain an analytical expression for the escape curve of the particle. This will be performed by leveraging the fact that a type I bifurcation (escape) is a particular case of the type II bifurcation analyzed in the previous section.

### 5.1. Maximum mechanism at $\nu = 0$

In this scenario, the LPT directly approaches the upper energy bound of the phase cylinder $\tilde{\xi} = 1$ at $\nu = 0$, without passing thought the saddle point at $\nu = \pi$. The critical excitation amplitude is obtained by substituting $\tilde{\xi} = 1$ into Eq. (25) as follows:

$$f_{m,0}(\Omega) = \frac{2}{\alpha}(1 - \beta\Omega) \tag{29}$$

### 5.2. Maximum mechanism at $\nu = \pi$

This escape scenario takes place when the saddle point vanishes, i.e. for $\Omega > \hat{\Omega}$. Then, the LPT directly approaches the upper energy bound of the phase cylinder $\tilde{\xi} = 1$ at $\nu = \pi$, in absence of a saddle point at $\nu = \pi$. The critical forcing amplitude is obtained from Eq. (27) by substituting $\tilde{\xi} = 1$ as follows:

$$f_{m,\pi}(\Omega) = -\frac{2}{\alpha}(1 - \beta\Omega) \tag{30}$$

## 5.3. Saddle mechanism

Since the SM is independent on $\tilde{\xi}$, it takes place as long as the saddle point exists, i.e. for $\Omega \in (0, \hat{\Omega}]$, as described by Eq. (22). For $\Omega > \hat{\Omega}$ the SM is overruled by the MM$\pi$. In other words, for excitation frequency of $\hat{\Omega}$ (Eq. (21)) there is a coexistence of both the SM and the MM$\pi$, as demonstrated in Fig. 9. Frequency $\hat{\Omega}$ and the corresponding excitation amplitude $\hat{f}$ are calculated in Eq. (21) and Eq. (23), respectively. Coexistence of both MM0 and the SM corresponds to the intersection of both mechanisms in Eq. (29) and Eq. (22), i.e. $(\Omega^*, f^*)$, as demonstrated on the phase portrait in Fig. 10. In this case, the following relation holds:

$$f_{m,0}(\Omega) = f_s(\Omega) \rightarrow \frac{4\beta^3}{27}\Omega^{*3} + \beta\Omega^* - 1 = 0 \rightarrow \Omega^* = \frac{3}{\beta}\left(\frac{\sqrt[3]{1+\sqrt{2}}}{2} - \frac{1}{2\sqrt[3]{1+\sqrt{2}}}\right) \approx 1.1362 \tag{31}$$

Substituting Eq. (31) into Eq. (22) yields the corresponding excitation amplitude as follows:

$$f^* = \frac{8\beta^3}{27\alpha}\Omega^3 = \frac{8}{\alpha}\left(\frac{\sqrt[3]{1+\sqrt{2}}}{2} - \frac{1}{2\sqrt[3]{1+\sqrt{2}}}\right)^3 \approx 0.2217 \tag{32}$$

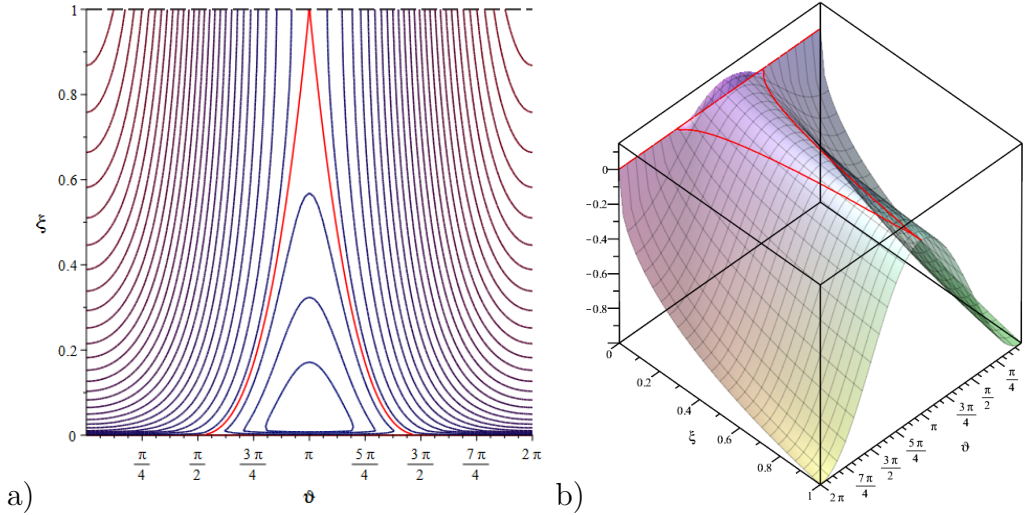

Figure 9: Coexistence of both the saddle mechanism and the maximum mechanism at $\nu = \pi$, for $\Omega = \hat{\Omega} = 1.9062$ and $f = \hat{f} = 1.0471$; a) 2D projection of $C(\nu, \xi)$ on the $(\nu, \xi)$ plane, b) 3D plot of $C(\nu, \xi)$.

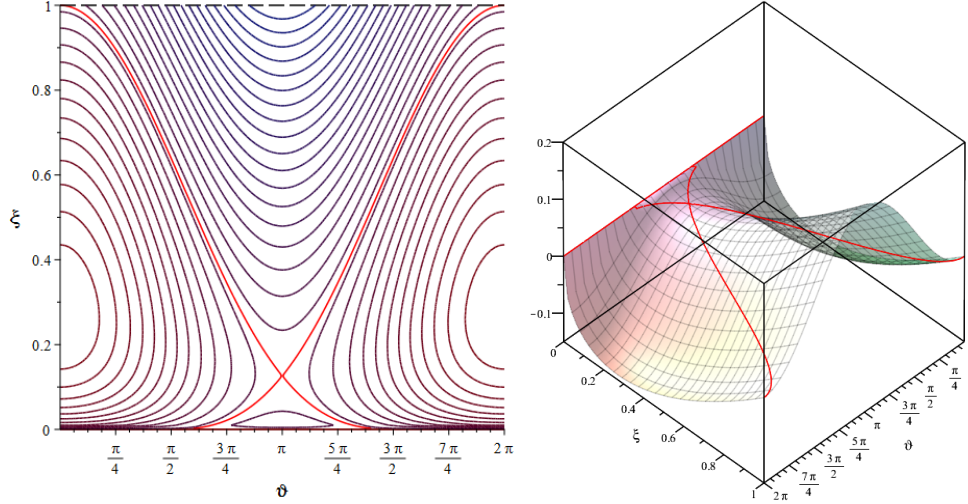

Figure 10: Coexistence of both the saddle mechanism and the maximum mechanism at $\nu = 0$, for $\Omega = \Omega^* = 1.1362$ and $f = f^* = 0.2217$; a) 2D projection of $C(\nu, \xi)$ on the $(\nu, \xi)$ plane, b) 3D plot of $C(\nu, \xi)$.

All three mechanisms correspond to three curves on the forcing parameters plane. The curve that is defined by $f(\Omega) = \max\{f_{m,0}(\Omega), f_s(\Omega), f_{m,\pi}(\Omega)\}$ corresponds to the escape envelope of the well, as shown in Fig.11. In other terms, for any excitation frequency, the upper branch among the triplet corresponds to the overruling escape mechanism that governs the transient escape process. Surprisingly, as one can see in Fig. 11, the universal property of a sharp minimum of the escape curve exists also in the current case, when the potential well lacks a linear term and the well is purely nonlinear. The right shift of the dip corresponds to the hardening effect of the quartic nonlinearity, i.e. the positive cubic term in the equation of motion (Eq. (2)).

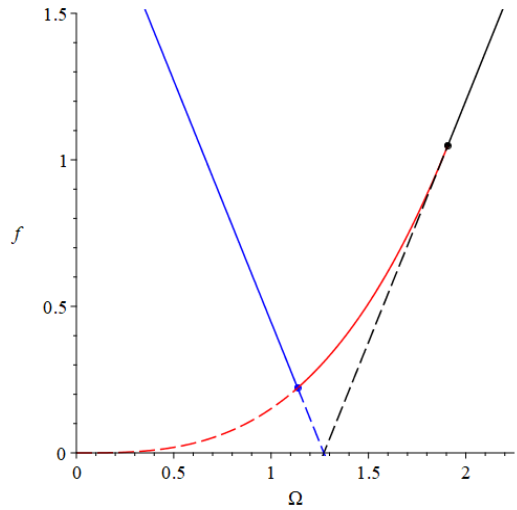

Figure 11: Escape envelope of a harmonically excited classical particle from a quartic potential well. The blue and black branches correspond to escape through the MM0 and the MM$\pi$, respectively. The red line corresponds to escape through the saddle mechanism. The coexistence of the MM0 and SM is denoted by a blue circle, and the coexistence of the MM$\pi$ and the SM is denoted by a black circle. A dashed line means that the dynamical mechanism is overruled by another mechanism.

## 6. Frequency response

Due to the importance of the response energy to the escape phenomenon and due to the engineering interest of this quantity, the energy-based frequency response of the particle is analyzed. The frequency response curve is defined as the intersection curve between the manifold $\tilde{\xi}(\Omega, f)$ and a plane that corresponds to a desired forcing amplitude $f$. The response curve is calculated based on Eq. (28) as shown in Fig. 12. The left branch and the right branch of each frequency response curve correspond to type II bifurcation through the MM0 and the MM$\pi$, respectively. The horizontal dashed line corresponds to escape from the well (type I bifurcation, $\tilde{\xi} = 1$), and the vertical dashed line corresponds to the sudden energy jump associated with a type II bifurcation through the SM. The intersection points between the left branch and the horizontal dashed line, and between the right branch and the vertical dashed line corresponds to type I bifurcation through the MM0 and MM$\pi$, respectively.

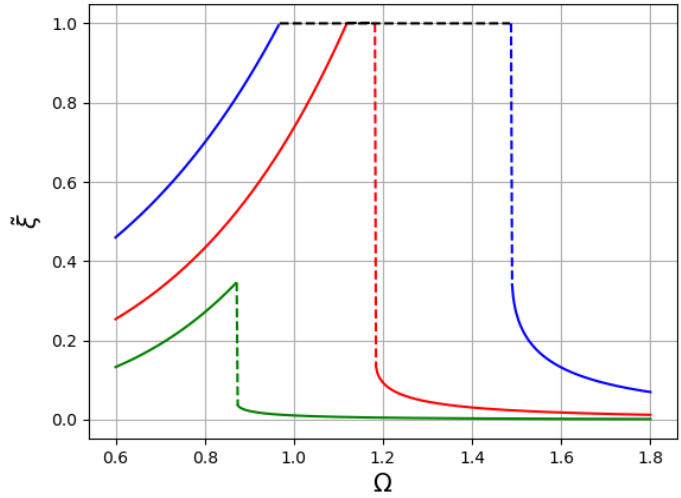

Figure 12: a) Frequency response curves for $f = 0.1, 0.25, 0.5$, colored in green, red, and blue, respectively. Vertical dashed lines correspond to non-smooth energy jumps associated with crossing the transition boundary that correspond to a bifurcation of type II through the saddle mechanism. Horizontal black dashed corresponds to escape (bifurcation of type I, $\tilde{\xi} = 1$).

## 7. Numerical verification

In this section, the analytical predictions of the escape envelope and frequency response curve are validated numerically. Since the analytical treatment is based on the assumption that the forcing frequency is in the vicinity $\Omega = 1$, we expect to have the validity of the approximate analysis near this frequency and degradation for significantly smaller/larger frequencies. Hence, we restrict the numerical analysis to the vicinity of the sharp dip of the escape curve, i.e. $(\Omega^*, f^*)$. Hence, the MM$\pi$ is not investigated herein and was mainly introduced for obtaining the inverse relation $\tilde{\xi}(\Omega, f)$ for $f < f_s(\Omega)$. All numerical results are obtained by integrating Eq. (2). Fig. 13-14 demonstrate escape through the MM0 and SM, respectively. Without loss of generality, those simulations support the bifurcation of type I, however similar simulations for type II bifurcation will have identical properties except the maximal transient energy reached ($\tilde{\xi} < 1$). In Fig. 13 one can see the gradual increase of the response displacement/energy until escape takes place. On the other hand, in Fig. 14 we demonstrate the characteristic sudden jump in the response energy associated with the SM.

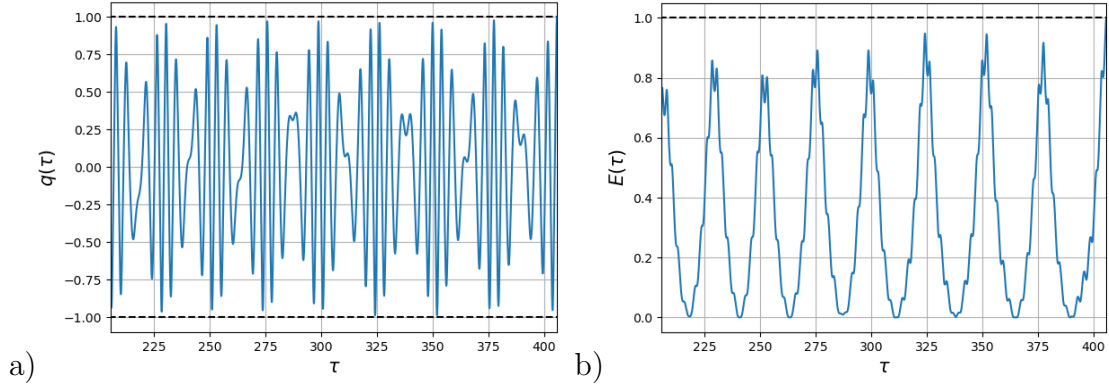

Figure 13: Bifurcation of type I ($\tilde{\xi} = 1$) through the maximum mechanism, for $\Omega = 1.115$ and $f = 0.219$, a) displacement response, b) energy response. Dashed black lines correspond to the bifurcation energy value, i.e. $\tilde{\xi} = 1$.

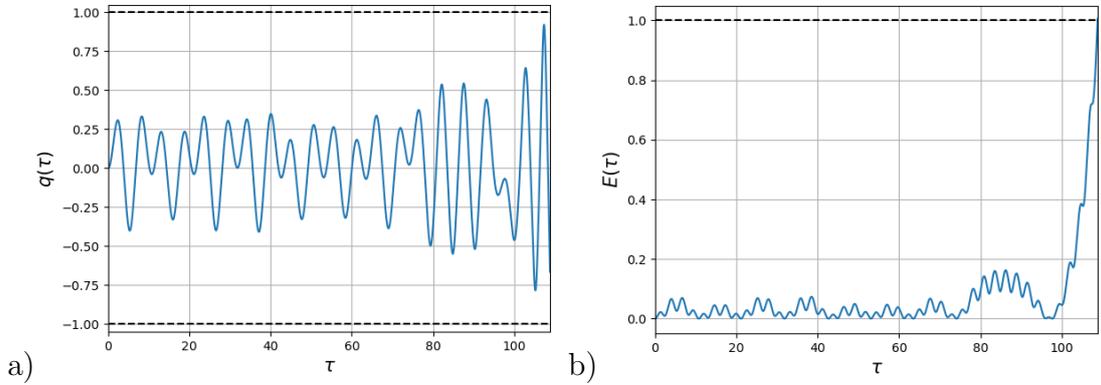

Figure 14: Bifurcation of type I (escape, $\tilde{\xi} = 1$) through the saddle mechanism, for $\Omega = 1.184$ i.e. $f = 0.25$, a) displacement response, b) energy response. Dashed black lines correspond to the bifurcation energy value, i.e. $\tilde{\xi} = 1$.

In Fig. 15(a) the analytical prediction of the escape curve is compared with numerical results. As one can and as expected, the analytical approximation is in good agreement with numerical results in the vicinity of the escape envelope's minimum (smaller $f$ values). As shown in Fig. 15(b), the numerical results are in good agreement with the predicted frequency response curve for $\Omega$ values that correspond to the bifurcation of type II through both the MM0 and the MM$\pi$. However, noisy results are observed on the branch associated with the MM$\pi$ near the occurrence of the SM. This stochastic nature stems from the sensitivity of the SM to small perturbation due to higher frequencies that are omitted in the analytical treatment. This sensitivity corresponds to the nature of the SM itself: a small divergence of the numerical results from the predicted LPT can cause the system to reach the saddle point and then to abruptly jump to a higher energy level. This chaotic-like nature of the escape through the SM becomes dominant also in the time required for an escape near the SM branch. The 'wiggly' look of the numerical results in Fig. 15(a) gives only a partial perspective on the noisy point-cloud of critical forcing amplitude values that are obtained for other terminal simulation times. This chaotic-like behavior and its stochastic properties are to be investigated in-depth in future studies. Moreover, the divergence between numerical and analytical results near the left branch of the frequency response curves in Fig. 15(b) increases with the forcing amplitude $f$. This finding corresponds to the increasing divergence between numerical results and analytical predictions in Fig. 15(a) as the excitation frequencies diverge from $\Omega^*$.

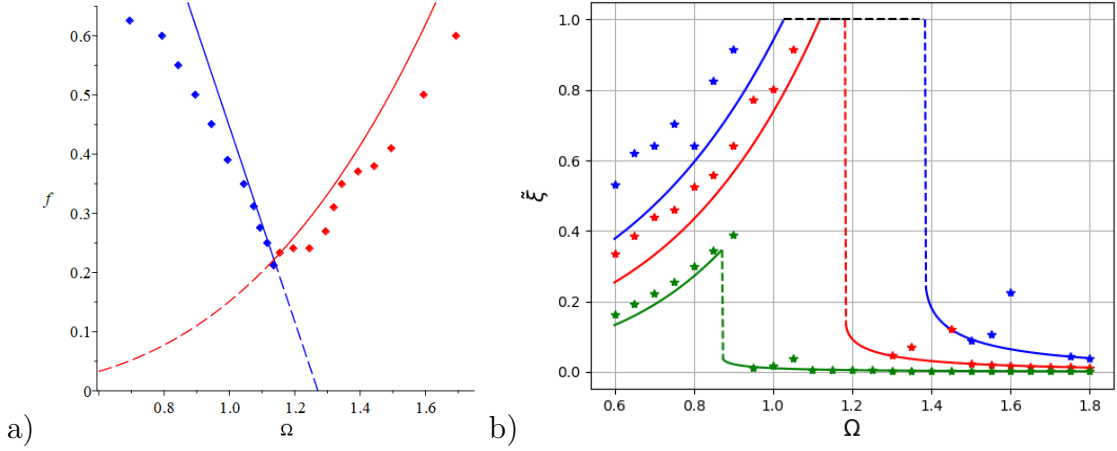

Figure 15: Numerical validation of the analytical results by integration of the equation on motion Eq. (2); a) escape curve (type I bifurcation, $\tilde{\xi} = 1$)- escape via MM0 and SM are marked by solid blue and red lines, respectively. A dashed line means that this escape mechanism is overruled by the other mechanism. Numerical simulations that correspond to escape through the MM and the SM are marked by blue and red points, respectively; b) frequency response curves for $f = 0.1, 0.25, 0.5$, colored in green, red, and blue, respectively. Vertical dashed lines correspond to non-smooth energy jumps associated with crossing the transition boundary that corresponds to a bifurcation of type II through the saddle mechanism. Horizontal black dashed line corresponds to escape (bifurcation of type I, $\tilde{\xi} = 1$). Numerical simulations for each case are marked in identical colors as their corresponding response curves.

## 8. Concluding remarks

In the current study, the transient escape problem of a harmonically-excited particle from a purely quartic potential well was investigated. This system can be observed as an equivalent model for physical and engineering systems with a cubic restoring force such as metal wires and springs, polymer structural components, and the cubic nonlinear energy sink. While previous works pointed out two underlying dynamical mechanisms, here three distinct mechanisms were identified and described analytically, giving a full understanding of the transient phenomena that govern the amplitude growth and escape. Two mechanisms are associated with a gradual increase of the particle's response amplitude and thus called maximum mechanisms, and the latter corresponds to a sudden increase in the response amplitude and called the saddle mechanism. Therefore, the latter is more potentially hazardous for structural elements and engineering systems but a more attractive regime for passive energy absorption systems. Additionally, a chaotic-like nature was identified in the critical parameters that lead to escape through the saddle mechanism and the time associated with the escape. This stochastic behavior is to be explored in-depth in future research. Due to the strong nonlinearity of the system, traditional perturbation-based methods are inapplicable for describing the system and thus canonical transformation to action-angle variables was utilized. The slow flow dynamics was described by a cylindrical reduced resonance manifold, and two types of bifurcations were defined: the former corresponds to escape from the well (maximal transient energy of one), and the latter to reaching a given energy level which is lower than one. The main result of this work is obtaining an analytical prediction for the escape envelope of the well, for the maximal transient energy level, and the frequency response curves of the particle for any given set of excitation parameters. All theoretical predictions are in good agreement with numerical results. Those findings and insights are to be considered in the design and optimization process of structural elements, mechanical components, and nonlinear passive energy absorbers with essential cubic nonlinearity in the stiffness/restoring force.

**Declarations**

**Funding** M. Farid has been supported by the Fulbright Program, the Israel Scholarship Education Foundation (ISEF), Jean De Gunzburg International Fellowship, the Israel Academy of Sciences and Humanities, the Yitzhak Shamir Postdoctoral Scholarship of the Israeli Ministry of Science and Technology, the PMRI – Peter Munk Research Institute - Technion, and the Israel Science Foundation Grant 1696/17.

**Availability of data and material** The data that supports the findings of this study is available from the author upon request.

# Appendix A. Detailed derivation of action-angle transformation

In this appendix, a detailed derivation of the action-angle transformation is described. The action variable is calculated by definition as follows:

$$I(E) = \frac{4}{2\pi}\int_0^{\sqrt[4]{E}} 2(E-q^4)dq = \frac{4}{\sqrt{2\pi}}E^{\frac{3}{4}}\int_0^1 \sqrt{1-u^4}du = \frac{4}{3\pi}\mathbf{K}\left(\frac{1}{\sqrt{2}}\right)E^{\frac{3}{4}} = \beta E^{\frac{3}{4}} \quad (A.1)$$

Here $u = q/\sqrt[4]{E}$ and $\mathbf{K}$ is the elliptic integral of the first kind. Now, the angle variable is derived according to Eq. (5):

$$\theta = \frac{\partial}{\partial I}\int_0^q \sqrt{2(E-U(x))}dx = \omega(E)\frac{\partial}{\partial E}\int_0^q \sqrt{2(E-U(x))}dx = \frac{\omega(E)}{\sqrt{2E}}\int_0^q \frac{1}{\sqrt{1-\frac{U(x)}{E}}}dx \quad (A.2)$$

Using Eq. (A.2), the angle variable is obtained as follows:

$$\theta = \frac{\omega(E)}{\sqrt{2}}E^{-\frac{1}{4}}\int_0^{q/\sqrt[4]{E}} \frac{1}{\sqrt{1-u^4}}du = \frac{\omega(E)}{\sqrt{2}}E^{-\frac{1}{4}}\mathbf{F}(u,i)\Big|_{u=0}^{q/\sqrt[4]{E}} = \frac{\omega(E)}{\sqrt{2}}E^{-\frac{1}{4}}\mathbf{F}(q/\sqrt[4]{E},i) \quad (A.3)$$

Here $i = \sqrt{-1}$, and $\mathbf{F}$ is the incomplete elliptic integral of the first kind. By inverting Eq. (A.3) we obtain the following expression for the displacement:

$$u(E,\theta) = \mathrm{sn}\left(\frac{\sqrt{2}\theta E^{\frac{1}{4}}}{\omega(E)}, i\right) \to q(E,\theta) = E^{\frac{1}{4}}\mathrm{sn}\left(\frac{\sqrt{2}\theta E^{\frac{1}{4}}}{\omega(E)}, i\right) \quad (A.4)$$

Fourier coefficients of the displacement $q(E,\theta)$ are obtained by the well-known nomal expansion of the elliptic function:

$$\mathrm{sn}(z,k) = \frac{2\pi}{\mathbf{K}(k)k}\sum_{n=0}^{\infty} \frac{Q^{n+1/2}\sin((2n+1)\zeta)}{1-Q^{2n+1}}, \quad Q = \exp\left(\frac{-\pi\mathbf{K}(k')}{\mathbf{K}(k)}\right), \zeta = \frac{\pi z}{2\mathbf{K}(k)} \quad (A.5)$$

Here $k' = \sqrt{1-k^2}$. Hence, the first term of the series is as follows:

$$\mathrm{sn}(z,k) \approx \frac{2\pi}{\mathbf{K}(k)k}\frac{\sqrt{Q(k)}}{1-Q(k)}\sin(\zeta) \quad (A.6)$$

The first order approximation of the displacement can be written as follows:

$$q(E,\theta) = \gamma(E)\sin(\theta) = \sum_{n=1}^{\infty} a_n\sin(n\theta), \quad \gamma(E) = \frac{2\pi\eta}{\mathbf{K}(i)(1+\eta^2)}E^{\frac{1}{4}} = \alpha E^{\frac{1}{4}} \quad (A.7)$$

Here $\eta = \sqrt{|Q(i)|} = e^{-\pi/2}$. The coefficient of the first term in the series is obtained as follows:

$$a_1(E) = \frac{1}{\pi}\int_{-\pi}^{\pi} \gamma(E)\sin(\theta)\sin(\theta)d\theta = \gamma(E) = \alpha E^{\frac{1}{4}} \quad (A.8)$$

# List of Abbreviations

| | |
|---|---|
| AA | action-angle (variables) |
| LPT | Limiting phase trajectory |
| MM | Maximum mechanism |
| MM0, MM$\pi$ | Maximum mechanism at $\nu = 0, \pi$, respectively |
| NES | nonlinear energy sink |
| PEA | Passive energy absorber |
| RM | Resonance manifold |
| SM | Saddle mechanism |

# List of Symbols

| | |
|---|---|
| $(\dot{\cdot}), (\cdot)'$ | Differentiation with respect to time scale $\tau$ and to averaged energy $\xi$, respectively |
| $\hat{\Omega}, \hat{f}$ | Excitation frequency and amplitude associated with the coexistence of the MM$\pi$ and the SM. |
| **F** | The incomplete elliptic integral of the first kind |
| **K** | The elliptic integral of the first kind |
| $\nu$ | Phase variable |
| $\omega$ | Response frequency of the particle |
| $\Omega^*, f^*$ | Excitation frequency and amplitude associated with the coexistence of the MM0 and the SM |
| $\tau$ | Non-dimensional time scale |
| $\mathrm{sn}(z, k)$ | The Jacobi elliptic function of module $k$ |
| $\tilde{\xi}$ | The maximal transient energy level |
| $\xi_s$ | The transient energy level associates with the saddle point of the RM |
| $a_k$ | The coefficient of the $k^{th}$ term in Fourier series of the solution $q(I, \theta)$ |
| $C(\nu, \xi)$ | The expression describes the resonance manifold |
| $E, \xi$ | Instantaneous and averaged energy of the particle |
| $f, \Omega$ | Non-dimensional forcing amplitude and frequency |
| $f_m, f_s$ | The critical forcing amplitudes associated with the occurrence of bifurcation through through maximum and saddle mechanisms, respectively |
| $H$ | Integral of motion/conservation law |
| $H_0$ | Initial conditions-related value of the integral of motion $H$ |
| $i$ | Unit imaginary number |
| $I, \theta$ | Action and angle variables |
| $J$ | Averaged action variable |
| $p$ | Momentum of the particle |
| $q$ | Non-dimensional displacement of the particle |
| $U(q)$ | Equivalent potential energy function |